# Inferring Questions from Programming Screenshots

Faiz Ahmed
*York University*
Toronto, ON, Canada
faiz5689@yorku.ca

Xuchen Tan
*York University*
Toronto, ON, Canada
swithin@yorku.ca

Folajinmi Adewole
*York University*
Toronto, ON, Canada
fj2005@my.yorku.ca

Suprakash Datta
*York University*
Toronto, ON, Canada
datta@yorku.ca

Maleknaz Nayebi
*York University*
Toronto, ON, Canada
mnayebi@yorku.ca

*Abstract*—The integration of generative AI into developer forums like Stack Overflow presents an opportunity to enhance problem-solving by allowing users to post screenshots of code or Integrated Development Environments (IDEs) instead of traditional text-based queries. This study evaluates the effectiveness of various large language models (LLMs)—specifically LLAMA, GEMINI, and GPT-4o in interpreting such visual inputs. We employ prompt engineering techniques, including in-context learning, chain-of-thought prompting, and few-shot learning, to assess each model's responsiveness and accuracy. Our findings show that while GPT-4o shows promising capabilities, achieving over 60% similarity to baseline questions for 51.75% of the tested images, challenges remain in obtaining consistent and accurate interpretations for more complex images. This research advances our understanding of the feasibility of using generative AI for image-centric problem-solving in developer communities, highlighting both the potential benefits and current limitations of this approach while envisioning a future where visual-based debugging copilot tools become a reality.

## I. INTRODUCTION

The increasing shift toward rich media communication across social media and online platforms raises the question of whether developers on Stack Overflow could solely use images, such as screenshots of code or their entire IDE, with little to no accompanying text, to convey their technical issues. Being mainly motivated by recent advancements in generative AI, which have made it possible for AI tools to analyze and interpret visual content more effectively, we explore how these developments could enable richer and potentially more intuitive interactions on platforms traditionally focused on text-based communication.

This trend toward rich media communication is evident across various social media platforms. As of April 2024, YouTube and Instagram remain among the most used platforms, with 83% and 47% of U.S. adults reporting usage, respectively [1]. Additionally, global users with Android devices spend an average of 31 hours and 32 minutes per month on TikTok, highlighting the growing appeal of visual content [2]. The rich media content plays a significant role in modern communication. For instance, a 2020 study on Stack Overflow and Bugzilla showed that the number of shared images on these platforms doubled over the past decade, demonstrating the increasing integration of visual media in technical discussions [3].

Our research is driven by developers' challenges in articulating technical questions effectively. Crafting precise and well-structured programming queries often demands considerable time and effort, frequently necessitating iterative refinements and prolonged discussions within developer forums. This process can be particularly daunting for users who may be apprehensive about community reception, adherence to platform guidelines, and potential impacts on their reputation, such as receiving negative feedback or having their posts flagged [4]. As such, many developers decide to use complementary multimedia, including screenshots, along with their written questions.

While current guidelines on Stack Overflow emphasize text-based, detailed questions to ensure clarity and searchability [5], [6], [7], the intuitive nature of sharing visual content, supported by advancements in generative AI [8], [9], [10], could change the dynamics of technical problem-solving [11]. This paper argues that leveraging generative AI could make processing and understanding images as easy as handling text [12], turning screenshots into viable inputs for community problem-solving and shifting the boundaries of how developers seek help and share knowledge.

As communication evolves from text to more visual formats, it becomes essential to assess the effectiveness of this shift in technical forums. Screenshots and visual media capture nuanced information quickly, making it easier for developers to communicate complex coding issues without lengthy descriptions. However, there are challenges related to accessibility, search engine indexing, and maintaining the readability standards that Stack Overflow upholds. This work benchmarks different generative AI models and prompting techniques to assess whether current advancements make this shift feasible, exploring how realistic it is and how far we are from achieving effective visual-based problem-solving in developer forums. In particular, we answer:

> How do various LLMs and different prompt engineering techniques perform in inferring questions given code and IDE screenshots?

## II. RELATED WORK

### A. Image Analysis for Software Tasks

With advancements in image processing, automatic analysis of images has become a tool in software engineering, particularly for testing purposes. Techniques have been developed to leverage image recognition for detecting user interface inconsistencies and improving the efficiency of software testing processes [13], [14], [15].

In addition to testing, image processing has been explored to facilitate developer communication and increase productivity on social coding platforms. Nayebi [3] investigated the role of visual content on platforms like Stack Overflow and Bugzilla, showing how images enhance communication and support collaborative work among developers. Later, Nayebi and Adams [16] reported the positive impact of visual elements on developer productivity in these settings. Other studies have also characterized visual content in developer environments, such as Jupyter Notebooks [17] and Stack Overflow [18]. Ahmed et al. [19] examined the use of image processing to identify duplicate questions on Stack Overflow, noting that while images can complement text, the improvements in efficiency were modest. Research has further focused on the frequent sharing of code screenshots on social coding platforms, with a particular emphasis on using machine learning models for accurate text retrieval from images [20], [21], [22].

*B. Image Analysis using LLMs*

LLMs have shown significant potential in image analysis across diverse fields by combining language and vision capabilities to perform complex multimodal tasks. Recent work integrates LLMs with vision models for applications such as image captioning, object recognition, and detailed description generation, supporting advancements in areas like healthcare, autonomous driving, and digital media [23], [24], [25]. Notably, models like CLIP [23] align images and text embeddings, enabling the categorization of images based on natural language descriptions. Similarly, multimodal models like BLIP [25] have expanded these capabilities by producing contextually relevant responses to image-based queries.

As such, prompt engineering was used to guide LLMs in analyzing images with various prompting techniques to improve accuracy and reasoning [26], [27]. Carefully crafted prompts have shown promise for specialized tasks, such as identifying objects in medical imaging or analyzing visual sentiment [28], [12], [24]. However, we could not identify any studies as such in software engineering [28], [29].

## III. METHOD

We applied various prompt engineering techniques to evaluate the performance of different Large Language Models (LLMs) in interpreting screenshot-based queries for developers' technical problem-solving. Specifically, we used in-context learning [30], chain-of-thought prompting [27], and few-shot learning [26] as our primary prompt strategies. In-context learning involves embedding relevant examples within the input prompt, allowing the model to draw from similar problem-solving patterns [30], [26]. Chain-of-thought prompting, which encourages the model to generate step-by-step reasoning, has proven effective for complex queries that require multi-step solutions [27]. Additionally, few-shot learning guides the models with minimal examples to balance adaptability and context comprehension.

We benchmarked three prominent LLMs in this evaluation: `LLaMA-3.2`[31], `Gemini-1.5-pro` [32], and `GPT-4o` [33]. We evaluated each model using various prompt engineering techniques to determine their responsiveness and accuracy in processing visual inputs from the programming environment. This comparative analysis identifies methods that improve interpretation and assesses the strengths and limitations of current generative AI technology in handling visual-based queries related to programming tasks. The findings provide insights into the feasibility of employing generative models as co-pilots [34], [35] for image-centric problem-solving on developer forums.

## IV. EXPERIMENT

*A. Data and Baseline*

We collected the most recent data from Stack Overflow using the Stack Exchange Data Explorer [36], focusing on questions with image attachments for October, starting October 1st to October $31^{st}$, 2024. To ensure the images present code or screenshots of IDEs, three authors—each with a minimum of four years of software development experience—independently reviewed the images. From an initial set of 4,310 Stack Overflow questions with some image attachments, we identified 2,610 posts that contained single image attachments to maintain consistency in our analysis. We then filtered posts containing relevant screenshots related to code or IDE-related, resulting in 152 Stack Overflow posts; after processing and filtering posts where some images were not readable by LLMs, we excluded them from the analysis. As a result, we evaluated 143 Stack Overflow questions, each with a single image of code or IDE screenshots.

We established a baseline using our dataset's original titles and bodies of questions as presented in Stack Overflow. This baseline enables us to evaluate the performance of LLMs in interpreting developers' visual inputs. We assessed the models' effectiveness in understanding and addressing screenshot-based queries by comparing LLM-generated responses to the original text-based content from Stack Overflow. This comparison helps determine how well LLMs can replicate or enhance the clarity and informativeness of traditional text-based questions and helps us discuss the potential for processing visual information in developer forums.

*B. Metrics*

We used a mixed-method approach in our experiment.

❶: We employed a vector-based embedding model to compute the *similarity between LLM-generated responses and the original questions*. We used `all-MiniLM-L6-v2`, a highly-rated and widely-used embedding model on HUGGING FACE [37]. We used this model to generate embeddings for both the original questions and LLM responses, and we calculated the similarity between these embeddings using cosine similarity [38].

❷ We used developers' perceptions to assess the practical applicability of our LLM-based approach. First, we examined the *relevance of the posted image and the accompanying*

*question* on Stack Overflow. Then, we measured both the *image's relevance to the LLM-generated question* and the *alignment between LLM-generated responses and the original question* on Stack Overflow. To guide these evaluations, we employed three key questions:

**Q1:** To what extent does the LLM response relate to the image? (Likert scale 0-10)
**Q2:** To what extent does the LLM response relate to the original Stack Overflow question? (Likert scale 0-10)
**Q3:** To what extent is the Stack Overflow question related to the image posted? (Likert scale 0-10)

We randomly selected 50 images for evaluation, with two experienced developers (minimum four years in software development) scoring each screenshot on Q1, Q2, and Q3. We calculated the average score per question for each screenshot and normalized the results to a scale from 0 to 1. We used this survey to facilitate [39]. The Cohen's kappa agreement among the annotators was 0.86, indicating the level of inter-rater reliability. Any discrepancies greater than one point between the two developers were addressed by initially having them discuss and resolve their differences whenever possible. For remaining disagreements, the first author of the paper acted as a moderator to reach a final consensus.

### C. Implementation details

We employed three LLMs `Gemini-1.5-Pro` [40], `Llama-3.2` [41], and `GPT-4o` [33]. `Gemini` and `GPT` are known for their superior multimodal processing capabilities and `Llama` is frequently used open source LLM. We standardized the generation parameters for all models to ensure consistency across experiments. We set the temperature ($t$) to zero to produce deterministic outputs and configured *top ˘p* to 0.8 and *top ˘k* to 40 to control the diversity of text generation. The maximum output length was capped at 2,048 tokens to maintain uniformity in response length. For implementation, we used the `VertexAI` Python SDK for interfacing with `Gemini`, the `Groq` client for `Llama`, and the `OpenAI` API for `GPT`. All models processed `base64`-encoded screenshot images we extracted from Stack Overflow posts. We evaluated each model using three prompt types, as detailed in Figure 1. For the few-shot learning approach, we selected two examples manually from the most upvoted questions in our dataset. [42], [43] All our data and code are available in our GitHub repository [44] for reproducibility.

## V. RESULTS

We discuss evaluation using metrics in Section IV-B.

### A. Embedding-Based Similarity Evaluation

We calculated similarity scores using embeddings from `all-MiniLM-L6-v2`, to evaluate the *semantic similarity between LLM-generated responses and original question* on Stack Overflow (①:). For a more granular analysis, we separately compared similarity scores for question titles, bodies, and combined inputs to observe potential variations in model performance on shorter versus longer text. We consistently observed higher similarity scores when combining the title and body of questions. Specifically, `GPT-4o` with in-context learning and `Gemini` with chain-of-thought prompting achieved the highest average score of $0.59$ for these combined inputs. Additionally, `GPT` using chain-of-thought prompting and `Gemini` with in-context learning performed only negligibly lower, with scores of $0.58$.

TABLE I: Average text similarity of LLMs generated queries with the baseline for different prompts (①:)

| Unit | Prompt Technique | LLAMA | GEMINI | GPT-4o |
|---|---|---|---|---|
| Question title | In-Context Learning | 0.21 | 0.47 | 0.46 |
| | Chain-of-Thought | 0.22 | 0.46 | 0.45 |
| | Few-Shot Learning | 0.04 | 0.37 | 0.44 |
| Question body | In-Context Learning | 0.27 | 0.52 | 0.52 |
| | Chain-of-Thought | 0.30 | 0.52 | 0.51 |
| | Few-Shot Learning | 0.05 | 0.44 | 0.49 |
| Combined title & body | In-Context Learning | 0.31 | 0.58 | *0.59\** |
| | Chain-of-Thought | 0.35 | *0.59\** | 0.58 |
| | Few-Shot Learning | 0.07 | 0.52 | 0.56 |

In contrast, `Llama` consistently yielded lower similarity scores across various prompting techniques, underperforming relative to both `GPT` and `Gemini`. Additionally, few-shot learning led

---

**In-Context Prompting Strategy**
**Task:** Generate a realistic Stack Overflow post about the following IDE/code screenshot. **Context:** You are a programmer experienced in various technology stacks, encountering an issue in a project. The screenshot shows the problem without textual content. Generate a post that: 1. Follows Stack Overflow guidelines 2. Includes relevant code/IDE context 3. Clearly states expected vs. actual behavior 4. Matches Stack Overflow format

**Few-Shot Learning Strategy**
**Task:** As an expert developer and Stack Overflow analyst, generate a question based on this screenshot, following these examples:
Example 1: [Screenshot 1] TITLE: Example Title 1 BODY: Example detailed description 1
Example 2: [Screenshot 2] TITLE: Example Title 2 BODY: Example detailed description 2
Requirements: 1. Clear and concise question 2. Highlight key error messages 3. Match example format

**Chain-of-Thought Strategy**
**Task:** Analyze the screenshot and create a Stack Overflow question by following these steps:
1. Initial Observation - Visible elements in the screenshot - IDE/tool identification - Error messages/indicators
2. Problem Identification - Main technical issue - Components involved - Issue type (configuration/syntax/runtime)
3. Context Building - Required technical background - Framework/language versions - Potential causes
4. Solution Attempts - Possible fixes tried - Relevant documentation - Troubleshooting steps
5. Question Formulation - Clear title - Technical details - Answerable format

**Output Format**
TITLE: Generated Title BODY: Generated Body

Fig. 1: Prompt used for inferring questions (Made a little short due to space constraint)

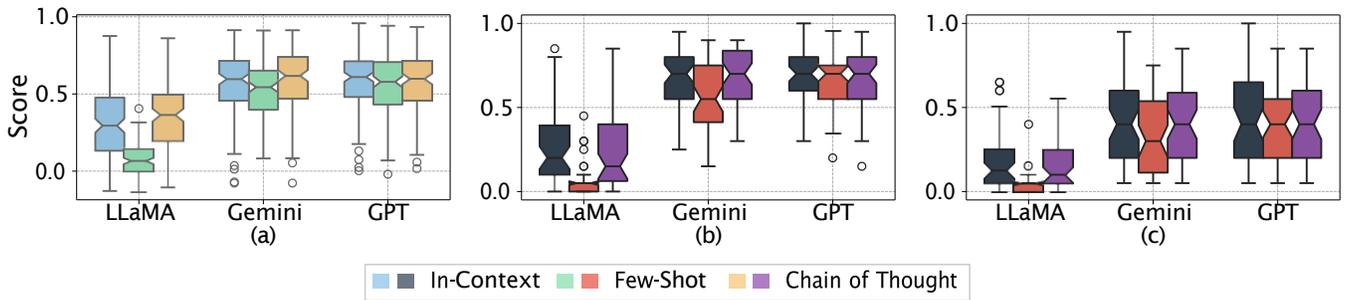

Fig. 2: (a) Embedding-based similarity for generated and original question (O:) (b) Relevance of generated question to screenshot by developers (�) and (c) Relevance of generated to the original question by developers (�)

to the lowest similarity scores, with averages ranging from $0.07$ to $0.56$ across models. This suggests that prompting style and the complexity of input texts (titles vs. bodies) play a significant role in the effectiveness of embedding-based similarity measures, yet the degree of similarity between the generated text, while promising, is still unclear. Overall, from Figure 2 and Table I, it is evident that both `GPT-4o` and `Gemini-1.5-pro` mostly maintained similarity scores predominantly above $0.4$, while `LLAMA-3.2` showed significantly lower and more variable performance with means of $\mu = 0.31, 0.07,$ and $0.36$.

### B. Developer-Assessed Usefulness

Two developers evaluated the relevance of LLM-generated questions to the screenshots (Q1) and their alignment with the original Stack Overflow questions (Q2). Our results, as shown in Figure 2 and summarized in Table II, demonstrate varying performance across different LLMs and prompting techniques.

For the relevance to screenshots (Q1), both `Gemini` and `GPT-4o` performed strongly, achieving scores of $0.67 - 0.68$ with in-context learning and chain-of-thought prompting. This consistent performance indicates their robust capability in interpreting visual content. `Llama`, in contrast, showed markedly lower relevance scores across all techniques, particularly struggling with few-shot learning, where it achieved only $0.06$, suggesting significant limitations in its ability to effectively interpret screenshot context.

For the relevance of LLM-generated questions relative to the original Stack Overflow question (Q2), all models demonstrated a decrease in alignment, particularly in few-shot learning. `GPT-4o` achieved the highest scores in this metric ($0.39 - 0.43$ across all prompt techniques), showing a moderate but noticeable alignment with the original question. `Gemini`'s relevance to the original question was slightly lower than `GPT-4o`'s, scoring around $0.34 - 0.41$. `Llama` again had the lowest performance, particularly in few-shot learning, where its relevance score dropped to $0.04$.

### VI. LIMITATIONS AND DISCUSSIONS

#### A. Relevance of Posted Images to Original Questions and Its Impact on LLM's Performance

In addition to evaluating the relevance of LLM-generated questions to both the screenshot and the original Stack Overflow question, we assessed the alignment between the posted Stack Overflow question and the associated image (Q3). Figure 3 illustrates this distribution, showing an average relevance score of $0.69$ while indicating that some screenshots have low relevance to the posted Stack Overflow question, suggesting inconsistencies in the alignment between visual content and the accompanying text-based query.

We further examined whether the relevance of the posted screenshot to the original question (Q3) correlates with the similarity between the LLM-generated question and the original question (Q2). Suppose the posted screenshot is only loosely related to the original question. In that case, it follows that the LLM-generated query—based solely on the screenshot—might also diverge from the original question. We identified a moderate correlation between these two factors (average of $0.53$), suggesting that while the LLM-generated question does partially reflect the original question's intent when the screenshot is relevant, this alignment is not consistently strong. This finding highlights a moderate performance and potential limitation in the LLM's ability to interpret questions raised in programming screenshots independently.

TABLE II: Average developer perceived similarity of LLMs generated questions with the baseline (�)

| Unit | Prompt Technique | Llama | Gemini | GPT-4o |
|---|---|---|---|---|
| **Q1: Relevance of LLM question to screenshot** | In-Context Learning | 0.28 | *0.68\** | 0.67 |
| | Chain-of-Thought | 0.26 | *0.68\** | 0.67 |
| | Few-Shot Learning | 0.06 | 0.56 | 0.65 |
| **Q2: Relevance of LLM question vs original** | In-Context Learning | 0.19 | 0.41 | *0.43\** |
| | Chain-of-Thought | 0.17 | 0.39 | 0.40 |
| | Few-Shot Learning | 0.04 | 0.34 | 0.39 |

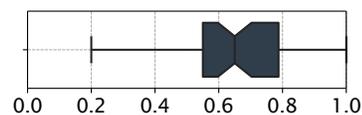

Fig. 3: Image relevance to questions on Stack Overflow (�)

*B. Characteristics of screenshot with best and worst LLM generated questions*

Our manual analysis of the ten best- and worst-performing images using `GPT-4o` with in-context prompting (the best-performing model from our experiment) revealed that optimal results were achieved with screenshots containing both error messages and code, where the error was explicitly visible. These screenshots often included relevant background information, such as the names of software or development tools in use. In some high-performing cases, users added annotations or highlights to pinpoint specific issues within the screenshot. Conversely, poorly performing screenshots typically displayed either code snippets or error messages alone, lacking contextual information from the IDE window. Additionally, screenshots with multiple, complex elements—such as numerous windows or menus—tended to confuse the model, further lowering performance.

This work builds upon the existing body of literature developed over the years [45], [19], [46], [47], [48], [49], [3], [50], [51], [52], [53], [54], [55], [56], [57], [58], [59], [60], [61], [62], [63], [64], [65], [66], [67], [68], [69], [70], [71], [72], [73], [74], [75], [76], [77], [78], [79].

*C. LLM Training and Data Exposure*

Considering that we used Stack Overflow's most recent data for this evaluation, it appears unlikely that the LLMs have been explicitly trained on this specific data; however, this cannot be definitively verified. This potential lack of specific data exposure could partially explain the weak correlations observed, as the models may not have context-specific training on Stack Overflow content or the particular images posted alongside questions. This ambiguity about the training data constitutes a potential threat to the validity of our findings. The dataset size ($n = 143$) constrained the study's scope, potentially limiting generalizability across diverse programming contexts.

*D. Choice of Metrics for Evaluation*

We selected embedding similarity as our primary evaluation metric because it effectively captures semantic meaning rather than relying on exact word matches, as is the case with traditional metrics such as BLEU and ROUGE. These conventional approaches are less suitable for evaluating open-ended question generation, as they depend on n-gram overlap and struggle to account for valid paraphrases or semantically equivalent questions [80], [81]. Nonetheless, to provide additional insights, we also computed BLEU and ROUGE scores, which are available in our supplementary materials [44]. The results from these exact-match metrics highlight significant variations, reflecting the natural diversity in developers' communication styles and varying levels of English fluency. For a qualitative assessment, we refer readers to the Q1 and Q2 evaluations in Section V-B. Additionally, we developed an interactive dashboard featuring six representative examples with detailed interpretations [82].

## VII. CAN SCREENSHOTS REPLACE TEXT QUESTIONS?

In this study, we investigated whether screenshots alone can effectively serve as inputs for technical problem-solving on Stack Overflow, focusing on how well LLMs can identify and generate relevant questions based solely on visual content. Using `Gemini-1.5-Pro`, `Llama-3.2`, and `GPT-4o` with various prompt techniques, we observed that LLM performance is significantly influenced by the relevance and clarity of the screenshots. Our evaluation demonstrated that state-of-the-art LLMs, particularly `Gemini` and `GPT-4o`, show promising capabilities in interpreting screenshot-based programming queries, with embedding-based similarity scores reaching up to $0.59$ and perceived relevance scores from developers reaching up to $0.69$. We observed that our best-performing model, `GPT-4o`, generated questions with over 60% similarity to the baseline (the original posted question) for 51.75% of the images. Upon further investigation, we found that GPT-4o's ability to accurately generate questions was strongly influenced by the content characteristics of the screenshots. Specifically, the presence of multiple windows or mixed content often led to confusion, reducing alignment with the original question and diluting the problem context. In contrast, `GPT-4o` performed best when screenshots displayed clear, singular content—such as code snippets, error messages, or user-created annotations—with in-context learning yielding the highest relevance scores in these cases. The potential for LLMs to support richer interactions with continued improvements in multimodal understanding is promising, yet the study highlights limitations when visual inputs are complex or ambiguous and we foresee that with further training and fine tuning of LLMs on the programming tasks and screenshots, copilots solely relying on images of IDE screens can be used for debugging programs and answering questions.

REFERENCES

[1] Pew Research Center, "Americans' social media use," 2024. Accessed: 2024-11-05.
[2] Hootsuite, "53 social media statistics to inform your 2024 social strategy," 2024. Accessed: 2024-11-05.
[3] M. Nayebi, "Eye of the mind: Image processing for social coding," in *Proceedings of the ACM/IEEE 42nd International Conference on Software Engineering: New Ideas and Emerging Results*, pp. 49–52, 2020.
[4] L. Park, "Research update: Improving the question-asking experience." https://stackoverflow.blog/2019/09/26/research-update-improving-the-question-asking-experience/, 2019. Accessed: 2024-12-15.
[5] batFINGER, "Policy on posting code / error message etc as images," 2021. Accessed: 2024-11-05.
[6] C. Gray, "Why should i not upload images of code/data/errors?," 2015. Accessed: 2024-11-05.
[7] apaul, "Discourage screenshots of code and/or errors," 2015. Accessed: 2024-11-05.
[8] R. Rombach, A. Blattmann, D. Lorenz, P. Esser, and B. Ommer, "High-resolution image synthesis with latent diffusion models," in *Proceedings of the IEEE/CVF Conference on Computer Vision and Pattern Recognition*, pp. 10684–10695, 2022.
[9] A. Ramesh, M. Pavlov, G. Goh, S. Gray, C. Voss, A. Radford, M. Chen, and I. Sutskever, "Zero-shot text-to-image generation," *arXiv preprint arXiv:2102.12092*, 2021.
[10] A. Razavi, A. van den Oord, and O. Vinyals, "Generating diverse high-fidelity images with vq-vae-2," in *Advances in Neural Information Processing Systems*, vol. 32, 2019.

[11] J. Wang, Z. Yang, X. Hu, L. Li, K. Lin, Z. Gan, Z. Liu, C. Liu, and L. Wang, "Git: A generative image-to-text transformer for vision and language," *arXiv preprint arXiv:2205.14100*, 2022.

[12] C. Zhang, W. Wang, P. Pangaro, N. Martelaro, and D. Byrne, "Generative image ai using design sketches as input: Opportunities and challenges," in *Proceedings of the 15th Conference on Creativity and Cognition*, pp. 254–261, 2023.

[13] T.-H. Chang, T. Yeh, and R. C. Miller, "Gui testing using computer vision," in *Proceedings of the SIGCHI Conference on Human Factors in Computing Systems*, pp. 1535–1544, 2010.

[14] T. D. White, G. Fraser, and G. J. Brown, "Improving random gui testing with image-based widget detection," in *Proceedings of the 28th ACM SIGSOFT international symposium on software testing and analysis*, pp. 307–317, 2019.

[15] E. Borjesson and R. Feldt, "Automated system testing using visual gui testing tools: A comparative study in industry," in *2012 IEEE Fifth International Conference on Software Testing, Verification and Validation*, pp. 350–359, IEEE, 2012.

[16] M. Nayebi and B. Adams, "Image-based communication on social coding platforms," *Journal of Software: Evolution and Process*, vol. 36, no. 5, p. e2609, 2024.

[17] V. Agrawal, Y.-H. Lin, and J. Cheng, "Understanding the characteristics of visual contents in open source issue discussions: a case study of jupyter notebook," in *Proceedings of the 26th International Conference on Evaluation and Assessment in Software Engineering*, pp. 249–254, 2022.

[18] H. Kuramoto, D. Wang, M. Kondo, Y. Kashiwa, Y. Kamei, and N. Ubayashi, "Understanding the characteristics and the role of visual issue reports," *Empirical Software Engineering*, vol. 29, no. 4, p. 89, 2024.

[19] F. Ahmed, S. Datta, and M. Nayebi, "Negative results of image processing for identifying duplicate questions on stack overflow," *arXiv preprint arXiv:2407.05523*, 2024.

[20] T. Beltramelli, "pix2code: Generating code from a graphical user interface screenshot," in *Proceedings of the ACM SIGCHI symposium on engineering interactive computing systems*, pp. 1–6, 2018.

[21] L. Bao, Z. Xing, X. Xia, D. Lo, M. Wu, and X. Yang, "psc2code: Denoising code extraction from programming screencasts," *ACM Transactions on Software Engineering and Methodology (TOSEM)*, vol. 29, no. 3, pp. 1–38, 2020.

[22] L. H. Li, M. Yatskar, D. Yin, C.-J. Hsieh, and K.-W. Chang, "Visualbert: A simple and performant baseline for vision and language," *arXiv preprint arXiv:1908.03557*, 2019.

[23] A. Radford, J. W. Kim, C. Hallacy, A. Ramesh, G. Goh, S. Agarwal, G. Sastry, A. Askell, P. Mishkin, J. Clark, *et al.*, "Learning transferable visual models from natural language supervision," in *International conference on machine learning*, pp. 8748–8763, PMLR, 2021.

[24] C. Jia, Y. Yang, Y. Xia, Y.-T. Chen, Z. Parekh, H. Pham, Q. Le, Y.-H. Sung, Z. Li, and T. Duerig, "Scaling up visual and vision-language representation learning with noisy text supervision," in *International conference on machine learning*, pp. 4904–4916, PMLR, 2021.

[25] J. Li, D. Li, C. Xiong, and S. Hoi, "Blip: Bootstrapping language-image pre-training for unified vision-language understanding and generation," in *International conference on machine learning*, pp. 12888–12900, PMLR, 2022.

[26] T. B. Brown, B. Mann, N. Ryder, M. Subbiah, J. D. Kaplan, P. Dhariwal, A. Neelakantan, P. Shyam, G. Sastry, A. Askell, *et al.*, "Language models are few-shot learners," *Advances in Neural Information Processing Systems*, vol. 33, pp. 1877–1901, 2020.

[27] J. Wei, X. Wang, D. Schuurmans, M. Bosma, B. Ichter, F. Xia, E. H. Chi, Q. V. Le, and D. Zhou, "Chain-of-thought prompting elicits reasoning in large language models," *arXiv preprint arXiv:2201.11903*, 2022.

[28] J. Wang, Z. Liu, L. Zhao, Z. Wu, C. Ma, S. Yu, H. Dai, Q. Yang, Y. Liu, S. Zhang, *et al.*, "Review of large vision models and visual prompt engineering," *Meta-Radiology*, p. 100047, 2023.

[29] J. Gu, Z. Han, S. Chen, A. Beirami, B. He, G. Zhang, R. Liao, Y. Qin, V. Tresp, and P. Torr, "A systematic survey of prompt engineering on vision-language foundation models," *arXiv preprint arXiv:2307.12980*, 2023.

[30] Q. Dong, L. Li, D. Dai, C. Zheng, J. Ma, R. Li, H. Xia, J. Xu, Z. Wu, T. Liu, *et al.*, "A survey on in-context learning," *arXiv preprint arXiv:2301.00234*, 2022.

[31] H. Touvron, T. Lavril, G. Izacard, X. Martinet, M.-A. Lachaux, T. Lacroix, B. Rozière, N. Goyal, E. Hambro, F. Azhar, *et al.*, "Llama: Open and efficient foundation language models," *arXiv preprint arXiv:2302.13971*, 2023.

[32] G. Team, R. Anil, S. Borgeaud, J.-B. Alayrac, J. Yu, R. Soricut, J. Schalkwyk, A. M. Dai, A. Hauth, K. Millican, *et al.*, "Gemini: a family of highly capable multimodal models," *arXiv preprint arXiv:2312.11805*, 2023.

[33] A. Hurst, A. Lerer, A. P. Goucher, A. Perelman, A. Ramesh, A. Clark, A. Ostrow, A. Welihinda, A. Hayes, A. Radford, *et al.*, "Gpt-4o system card," *arXiv preprint arXiv:2410.21276*, 2024.

[34] A. E. Hassan, G. A. Oliva, D. Lin, B. Chen, Z. Ming, *et al.*, "Rethinking software engineering in the foundation model era: From task-driven ai copilots to goal-driven ai pair programmers," *arXiv preprint arXiv:2404.10225*, 2024.

[35] A. Nguyen-Duc, B. Cabrero-Daniel, A. Przybylek, C. Arora, D. Khanna, T. Herda, U. Rafiq, J. Melegati, E. Guerra, K.-K. Kemell, *et al.*, "Generative artificial intelligence for software engineering–a research agenda," *arXiv preprint arXiv:2310.18648*, 2023.

[36] S. Exchange, "Stack exchange data explorer." https://data.stackexchange.com/. Accessed: 2024-11-07.

[37] Hugging Face, "model card." https://huggingface.co/sentence-transformers/all-MiniLM-L6-v2, 2024. Taken from the official website of Hugging Face.

[38] F. Rahutomo, T. Kitasuka, M. Aritsugi, *et al.*, "Semantic cosine similarity," in *The 7th international student conference on advanced science and technology ICAST*, vol. 4, p. 1, University of Seoul South Korea, 2012.

[39] S. App, "Question inferring tool." https://question-inferring.streamlit.app/, 2024. Accessed: 2024-11-05.

[40] G. AI, "Gemini 1.5 pro model card," 2024. Accessed: 2024-11-10.

[41] M. AI, "Llama 3.2 model card," 2024. Accessed: 2024-11-10.

[42] S. Overflow, "Few-shot example 1." https://stackoverflow.com/questions/79044080/is-getenv-s-not-part-of-cstdlib/, 2024. Accessed: 2024-11-05.

[43] S. Overflow, "Few-shot example 2." https://stackoverflow.com/questions/79084406/trying-to-stack-2-columns-into-one-excel, 2024. Accessed: 2024-11-05.

[44] Github, "Github repository." https://github.com/Research-Purpose/MSR-2025, 2024. Accessed: 2024-11-05.

[45] U. A. Koana, Q. H. Le, S. Raman, C. Carlson, F. Chew, and M. Nayebi, "Examining ownership models in software teams," *Empirical Software Engineering*, vol. 29, no. 6, pp. 1–43, 2024.

[46] S. G. Saroar, W. Ahmed, E. Onagh, and M. Nayebi, "Github marketplace for automation and innovation in software production," *Information and Software Technology*, vol. 175, p. 107522, 2024.

[47] S. G. Saroar and M. Nayebi, "Developers' perception of github actions: A survey analysis," in *Proceedings of the 27th International Conference on Evaluation and Assessment in Software Engineering*, pp. 121–130, 2023.

[48] U. A. Koana, F. Chew, C. Carlson, and M. Nayebi, "Ownership in the hands of accountability at brightsquid: A case study and a developer survey," in *Proceedings of the 31st ACM Joint European Software Engineering Conference and Symposium on the Foundations of Software Engineering*, pp. 2008–2019, 2023.

[49] M. Nayebi, B. Adams, and G. Ruhe, "Release practices for mobile apps– what do users and developers think?," in *2016 ieee 23rd international conference on software analysis, evolution, and reengineering (saner)*, vol. 1, pp. 552–562, IEEE, 2016.

[50] Y. Hashemi, M. Nayebi, and G. Antoniol, "Documentation of machine learning software," in *2020 IEEE 27th International Conference on Software Analysis, Evolution and Reengineering (SANER)*, pp. 666–667, IEEE, 2020.

[51] M. Nayebi, L. Dicke, R. Ittyipe, C. Carlson, and G. Ruhe, "Essmart way to manage customer requests," *Empirical Software Engineering*, vol. 24, pp. 3755–3789, 2019.

[52] M. Nayebi, G. Ruhe, and T. Zimmermann, "Mining treatment-outcome constructs from sequential software engineering data," *IEEE Transactions on Software Engineering*, vol. 47, no. 2, pp. 393–411, 2019.

[53] M. Nayebi, *Analytical Release Management for Mobile Apps*. PhD thesis, PhD thesis, University of Calgary, 2018.

[54] M. Nayebi, "Data driven requirements engineering: Implications for the community," in *2018 IEEE 26th International Requirements Engineering Conference (RE)*, pp. 439–441, IEEE, 2018.

[55] S. G. Saroar, W. Ahmed, E. Onagh, and M. Nayebi, "Github marketplace: Driving automation and fostering innovation in software de-

velopment," in *2025 IEEE 32nd International Conference on Software Analysis, Evolution, and Reengineering (SANER)*, Journal First, 2025.
[56] F. Pepe, C. Farkas, M. Nayebi, G. Antoniol, and M. Di Penta, "How do papers make into machine learning frameworks: A preliminary study on tensorflow," in *33rd IEEE/ACM International Conference on Program Comprehension (ICPC 2025)*, 2025.
[57] F. Ahmed, X. Tan, O. Adewole, and M. Nayebi, "Inferring questions from programming screenshots," in *22nd International Conference on Mining Software Repositories (MSR)*, 2025.
[58] M. Nayebi, K. Kuznetsov, A. Zeller, and G. Ruhe, "Recommending and release planning of user-driven functionality deletion for mobile apps," *Requirements Engineering*, vol. 29, no. 4, pp. 459–480, 2024.
[59] M. Nayebi, K. Kuznetsov, A. Zeller, and G. Ruhe, "User driven functionality deletion for mobile apps," in *2023 IEEE 31st International Requirements Engineering Conference (RE)*, pp. 6–16, IEEE, 2023.
[60] M. Nayebi, H. Cho, and G. Ruhe, "App store mining is not enough for app improvement," *Empirical Software Engineering*, vol. 23, pp. 2764–2794, 2018.
[61] M. Nayebi, M. Marbouti, R. Quapp, F. Maurer, and G. Ruhe, "Crowdsourced exploration of mobile app features: A case study of the fort mcmurray wildfire," in *2017 IEEE/ACM 39th International Conference on Software Engineering: Software Engineering in Society Track (ICSE-SEIS)*, pp. 57–66, IEEE, 2017.
[62] M. Nayebi and G. Ruhe, "Analytical product release planning," in *The art and science of analyzing software data*, pp. 555–589, Elsevier, 2015.
[63] M. Nayebi, H. Farahi, and G. Ruhe, "Which version should be released to app store?," in *2017 ACM/IEEE International Symposium on Empirical Software Engineering and Measurement (ESEM)*, pp. 324–333, IEEE, 2017.
[64] M. Nayebi, Y. Cai, R. Kazman, G. Ruhe, Q. Feng, C. Carlson, and F. Chew, "A longitudinal study of identifying and paying down architecture debt," in *2019 IEEE/ACM 41st International Conference on Software Engineering: Software Engineering in Practice (ICSE-SEIP)*, pp. 171–180, IEEE, 2019.
[65] M. Nayebi, H. Cho, H. Farrahi, and G. Ruhe, "App store mining is not enough," in *2017 IEEE/ACM 39th International Conference on Software Engineering Companion (ICSE-C)*, pp. 152–154, IEEE, 2017.
[66] M. Nayebi and G. Ruhe, "Asymmetric release planning: Compromising satisfaction against dissatisfaction," *IEEE Transactions on Software Engineering*, vol. 45, no. 9, pp. 839–857, 2018.
[67] W. Maalej, M. Nayebi, and G. Ruhe, "Data-driven requirements engineering-an update," in *2019 IEEE/ACM 41st International Conference on Software Engineering: Software Engineering in Practice (ICSE-SEIP)*, pp. 289–290, IEEE, 2019.
[68] M. Nayebi, K. Kuznetsov, P. Chen, A. Zeller, and G. Ruhe, "Anatomy of functionality deletion: an exploratory study on mobile apps," in *Proceedings of the 15th International Conference on Mining Software Repositories*, pp. 243–253, 2018.
[69] M. Nayebi and G. Ruhe, "Optimized functionality for super mobile apps," in *2017 IEEE 25th international requirements engineering conference (RE)*, pp. 388–393, IEEE, 2017.
[70] G. Ruhe, M. Nayebi, and C. Ebert, "The vision: Requirements engineering in society," in *2017 IEEE 25th International Requirements Engineering Conference (RE)*, pp. 478–479, IEEE, 2017.
[71] M. Nayebi, G. Ruhe, R. C. Mota, and M. Mufti, "Analytics for software project management–where are we and where do we go?," in *2015 30th IEEE/ACM International Conference on Automated Software Engineering Workshop (ASEW)*, pp. 18–21, IEEE, 2015.
[72] M. Nayebi and G. Ruhe, "An open innovation approach in support of product release decisions," in *Proceedings of the 7th International Workshop on Cooperative and Human Aspects of Software Engineering*, pp. 64–71, 2014.
[73] M. Nayebi, H. Farrahi, and G. Ruhe, "Analysis of marketed versus not-marketed mobile app releases," in *Proceedings of the 4th International Workshop on Release Engineering*, pp. 1–4, 2016.
[74] M. Nayebi, S. J. Kabeer, G. Ruhe, C. Carlson, and F. Chew, "Hybrid labels are the new measure!," *IEEE Software*, vol. 35, no. 1, pp. 54–57, 2017.
[75] G. Ruhe and M. Nayebi, "What counts is decisions, not numbers—toward an analytics design sheet," in *Perspectives on Data Science for Software Engineering*, pp. 111–114, Elsevier, 2016.
[76] M. Nayebi and G. Ruhe, "Analytical open innovation for value-optimized service portfolio planning," in *Software Business. Towards Continuous Value Delivery: 5th International Conference, ICSOB 2014, Paphos, Cyprus, June 16-18, 2014. Proceedings 5*, pp. 273–288, Springer, 2014.
[77] M. Nayebi, H. Farrahi, A. Lee, H. Cho, and G. Ruhe, "More insight from being more focused: analysis of clustered market apps," in *Proceedings of the International Workshop on App Market Analytics*, pp. 30–36, 2016.
[78] S. J. Kabeer, M. Nayebi, G. Ruhe, C. Carlson, and F. Chew, "Predicting the vector impact of change-an industrial case study at brightsquid," in *2017 ACM/IEEE International Symposium on Empirical Software Engineering and Measurement (ESEM)*, pp. 131–140, IEEE, 2017.
[79] M. Nayebi and G. Ruhe, "Trade-off service portfolio planning–a case study on mining the android app market," tech. rep., PeerJ PrePrints, 2015.
[80] E. Kamalloo, N. Dziri, C. L. Clarke, and D. Rafiei, "Evaluating open-domain question answering in the era of large language models," *arXiv preprint arXiv:2305.06984*, 2023.
[81] M. M. Bhat, R. Meng, Y. Liu, Y. Zhou, and S. Yavuz, "Investigating answerability of llms for long-form question answering," *arXiv preprint arXiv:2309.08210*, 2023.
[82] "Example analysis dashboard." https://example-analysis-msr2025.streamlit.app/, 2024. Accessed: 2024-01-31.